\documentclass[%
preprint,
amsmath,amssymb,
aps,
]{revtex4-1}

\usepackage{physics}
\usepackage{amsmath,amssymb}
\usepackage{mathtools}
\usepackage{graphicx}% Include figure files
\usepackage{dcolumn}% Align table columns on decimal point
\usepackage{bm}% bold math
\usepackage{color,soul}
\usepackage{array}
\usepackage{amsthm}
\usepackage{subcaption}
\usepackage{algorithm}
\usepackage{algpseudocode}
\usepackage{wrapfig}
\usepackage{qcircuit}
\usepackage{xcolor}
\usepackage{chngcntr}
\usepackage{setspace}
\usepackage[colorlinks,linkcolor=blue,anchorcolor=red,citecolor=red]{hyperref}
\usepackage[capitalize]{cleveref}
\newcounter{algsection}
\counterwithin{algorithm}{algsection}
\renewcommand\thealgorithm{%
\ifnum \value{algsection}>0%
{\arabic{algorithm}.\arabic{algsection}}%
\else
{\arabic{algorithm}}%
\fi}
\DeclarePairedDelimiter{\ceil}{\lceil}{\rceil}

\algnewcommand\algorithmicswitch{\textbf{switch}}
\algnewcommand\algorithmiccase{\textbf{case}}
\algnewcommand\algorithmicassert{\texttt{assert}}
\algdef{SE}[SWITCH]{Switch}{EndSwitch}[1]{\algorithmicswitch\ #1\ \algorithmicdo}{\algorithmicend\ \algorithmicswitch}%
\algdef{SE}[CASE]{Case}{EndCase}[1]{\algorithmiccase\ #1}{\algorithmicend\ \algorithmiccase}%
\algtext*{EndSwitch}%
\algtext*{EndCase}%
\algnewcommand{\IIf}[1]{\State\algorithmicif\ #1\ \algorithmicthen}
\algnewcommand{\EndIIf}{\unskip\ \algorithmicend\ \algorithmicif}

\begin{document}
	
	\title{Quantum walk-based vehicle routing optimisation}
	
	\author{T. Bennett}
% 	\email{20940505@student.uwa.edu.au}
	\affiliation{Department of Physics, The University of Western Australia, Perth, Australia}
	\author{E. Matwiejew}
% 	\email{edric.matwiejew@research.uwa.edu.au}
	\affiliation{Department of Physics, The University of Western Australia, Perth, Australia}
	\author{S. Marsh}
% 	\email{samuel.marsh@research.uwa.edu.au}
	\affiliation{Department of Physics, The University of Western Australia, Perth, Australia}
	\author{J. B. Wang}
	\email{jingbo.wang@uwa.edu.au}
	\affiliation{Department of Physics, The University of Western Australia, Perth, Australia}

	\date{\today}
	
	\begin{abstract}
	This paper demonstrates the applicability of the Quantum Walk-based Optimisation Algorithm (QWOA) to the Capacitated Vehicle Routing Problem (CVRP). Efficient algorithms are developed for the indexing and unindexing of the solution space and for implementing the required alternating phase-walk unitaries, which are the core components of QWOA. Results of numerical simulation demonstrate that the QWOA is capable of producing convergence to near-optimal solutions for a randomly generated 8 location CVRP. Preparation of the amplified quantum state in this example problem is demonstrated to produce high-quality solutions, which are more optimal than expected from classical random sampling of equivalent computational effort.  
	\end{abstract}
	
	\maketitle
	
	\section{Introduction}
    Quantum computers may offer a unique advantage when it comes to many combinatorial optimisation problems, where the search for an optimal solution becomes quickly infeasible due to solution spaces that scale exponentially with increasing problem size \cite{Aaronson2005}. While a classical computer's central processing unit is restricted to assessing the quality of solutions one after another, a quantum processor is capable of operating on the complete solution space at once via the principle of quantum superposition. However, this capability alone does not provide any significant utility. For example, consider a quantum system initialised in an equal superposition of states, one for each solution in the solution space. A measurement of the system is equally likely to collapse the system into any one of these states, which is equivalent to simply picking a solution at random. Where quantum superposition offers significant utility, is when it is combined with a suitable quantum algorithm. Such an algorithm should be capable of significantly increasing the probability of measuring the system in a state corresponding to the most, or one of the most optimal solutions.
    
    %A quantum computer is capable of operating on the complete solution space in quantum parallel, exploiting superposition and entanglement to boost the probability of finding higher-quality solutions using fewer iterations than is classically possible.
    
    One candidate is the Quantum Walk-based Optimisation Algorithm (QWOA)~\cite{Marsh2019,marsh2020combinatorial}. This algorithm makes use of alternating quality-dependent phase-shifts and continuous-time quantum walks over a circulant graph connecting all possible solutions in the solutions space. By tuning the applied phase-shifts and quantum walks via a set of classical parameters, it is hoped that quantum interference  will result in a concentration of probability density at states corresponding to high-quality solutions. This tuning process is carried out via a classical optimisation procedure which optimises for the expectation value of quality, as measured from the quantum circuit. Its effectiveness has been recently demonstrated in the context of portfolio optimisation problems~\cite{slate2020quantum}.
    
    In this paper, we show the applicability of the QWOA to the combinatorial optimisation problem of capacitated vehicle routing. The capacitated vehicle routing problem (CRVP) asks which route(s) should be taken by supply vehicles with a limited capacity in order to deliver products from a single depot to multiple locations, each requiring unique quantities of various products. Trips to and from the depot as well as between external locations are all characterised by route-dependent costs. The globally optimal solution is the route or set of routes that successfully delivers all required products whilst minimising the total delivery cost, subject to the further constraint that every vehicle must return to the depot upon completion of delivery. A general form of the VRP problem was first studied by \citet{dantzig1959truck} in 1959, and although algorithms have since been developed to solve smaller-scale problems exactly \cite{TOTH2002487}, the focus for larger scale problems has been on heuristic-based methods for finding near-optimal solutions. The VRP problem has clear real world significance because improving vehicle schedules, even by a tiny proportion, can lead to a large reduction in the transportation costs over time. Such problems have very recently been the subject of quantum algorithm development, in particular, using the quantum annealing approach \cite{harikrishnafkumar2020quantum, borowski2020new, SYRICHAS201752, 10.3389/fict.2019.00013}. 
    %Behera et al. attempted the QAOA approach to this problem, but concluded that their 
    %behera2020solving, 
    This paper, however, focuses on the application of QWOA to the Capacitated Vehicle Routing Problem (CVRP), an algorithm that operates within the gate-based model of quantum computation. The purpose of this paper is to demonstrate that the QWOA can be effectively applied to the CRVP, and to present and analyse the results of numerical simulations of the application of the algorithm to an example problem. 
    
    The structure of this paper is as follows: In Sec. II, the CVRP will be formally introduced, along with an illustrative example and a brief discussion of its solution space. In Sec. III, the QWOA will be introduced, including its theoretical framework and quantum circuit implementation. In Sec. IV, the CVRP will be shown to satisfy the necessary prerequisite features for application of the QWOA, including possessing efficient processes for computing the cardinality of the solution space, for indexing/unindexing of the solution space, and for computing solution qualities. Finally, in Sec. V, the numerical results of the simulated application of the QWOA to an example CVRP will be presented and analysed.
    
    \section{The Capacitated Vehicle Routing Problem}
    
    \subsection{Formal Definition}
    
    A capacitated vehicle routing problem consisting of $n$ delivery locations shall be referred to as a problem of size $n$. The delivery network for such a problem shall be characterised by a complete graph with $n+1$ vertices. The vertex representing the depot is labelled with a zero, and the delivery locations are represented by vertices labelled $1, 2, ..., n$. The number of packages required at each location are included in a package vector, $P$, of dimension $n$, containing non-negative integers, $P_i$, where $P_i$ is the number of packages required at location $i$ and $i = 1, 2, ..., n$. The costs associated with the trips between nodes are captured in a cost matrix, $C$. The cost matrix is square, $n+1$ dimensional, and element $C_{ij}$ is the cost of the trip from node $i$ to node $j$ with $i,j = 0, 1, ..., n$ and with $C_{ij}$ taking positive finite values. Since it makes no sense to talk of a trip from a node to itself, the leading diagonal of the cost matrix consists of zeros. If costs are equal in both directions for all trips between nodes, then the cost matrix will be symmetrical, however, this need not be the case. The vehicle capacity is represented by the natural number variable, $V$. No consideration is made for time, and any solution which minimises cost can be undertaken by a single vehicle. Thus, while particular solutions (delivery routes) may be undertaken with multiple vehicles, the number of vehicles does not, and need not, make an appearance in this formulation of the CVRP. 
    
    Given a particular instance of the CVRP, solving the problem reduces to finding a solution from the space of all possible solutions, $\mathbb{S}$, which minimises total cost. Or to put it more formally, for some instance of the CVRP, specified by a unique set of the parameters, $P$, $C$ and $V$, and with a quality function, $f(x)$, which returns the total cost of any given solution $x$, the problem reduces to finding an optimal solution $x^*$, such that
    \begin{equation}
        f(x^*) = \min \{ f(x) \mid x \in  \mathbb{S}\},
    \end{equation}
    The difficulty lies in the total number of possible solutions, $M$, which grows exponentially with increasing problem size $n$.
    
    \subsection{Illustrative Example}
    
\begin{figure}[H]
        \centering
        \includegraphics[width=0.85\columnwidth]{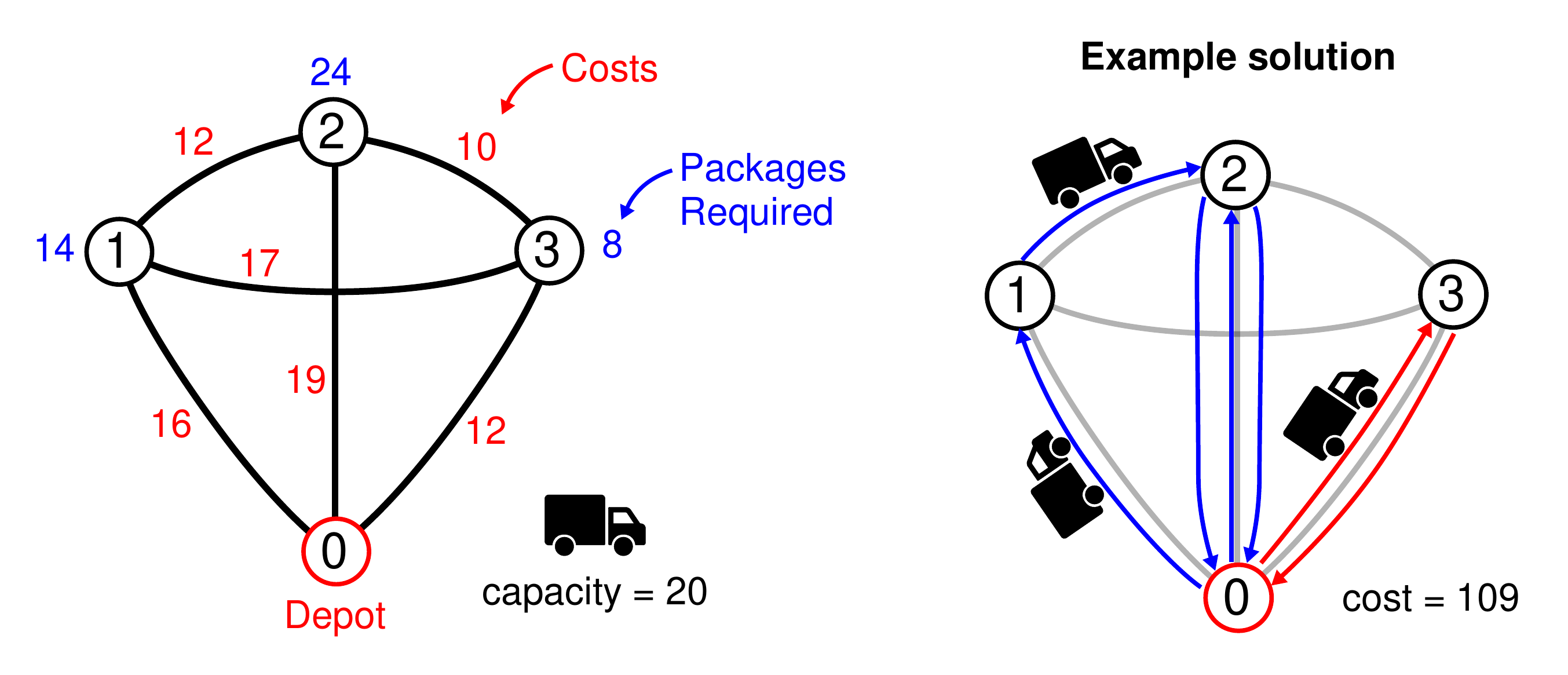}
        \caption{Shown on the left is an example of a CVRP of size $n=3$ and with vehicle capacity $V=20$. In blue are the required package numbers $P_i$, and each trip/edge has its cost $C_{ij}$ shown in red. On the right is the route for this particular problem corresponding to the solution (1, 2), (3).}
        \label{fig:illustrative_example}
    \end{figure}

    A CVRP is fully defined by the three aforementioned variables, $P$, $C$, and $V$. As an example, the following variables 
    \[ P = \left[ \begin{array}{c} 14 \\ 24 \\ 8\\ \end{array} \right], \;\; C= \left[ \begin{array}{cccc} 0 & 16 & 19 & 12 \\ 16 & 0 & 12 & 17 \\ 19 & 12 & 0 & 10 \\ 12 & 17 & 10 & 0 \\ \end{array} \right] , \;\; V = 20 \]
    fully specify a CVRP of size $n=3$, as illustrated in \cref{fig:illustrative_example}.
    
    The example solution shown in \cref{fig:illustrative_example} can be described as follows: A delivery is made to location 1, the leftover stock from this delivery is taken to location 2. The vehicle returns to the depot to restock before completing delivery of the remaining packages to location 2. Rather than taking the small number of leftover packages to location 3, the vehicle returns to the depot to restock before completing the round trip to location 3. Since the vehicle returned to the depot between locations 2 and 3, the solution is effectively split into two independent delivery groups/routes, one shown with blue arrows and the other with red. We can represent this solution as a set of subsets: \{(1, 2), (3)\}. The first subset corresponds to the route shown with blue arrows and the second corresponds to the route shown with red arrows. Note that the order of these subsets does not affect the quality of the solution, only the order of elements within each subset. 
    
    \subsection{Solution Space}
    
    The representation of the above example solution as an unordered set of ordered subsets can be extended to capture the entire solution space. Separation of the locations into independent delivery groups in all possible ways is akin to taking all possible partitions of the full set $\{1, 2, \ldots, n\}$. The order in which locations can be visited within each delivery group must also be taken into account. So for each partition, the full set of solutions it represents can be acquired by taking all combinations of the permutations of each subset. In other words, by first expanding each subset in the partition into a set containing all permutations of its elements, then by taking the Cartesian product of the resulting sets of permutations. The resulting solution space is the set of all possible partitions of the $n$ elements into nonempty and totally ordered subsets.
    
    \section{The Quantum Walk Optimisation Algorithm}
    
    \subsection{Theoretical Framework}

    The Quantum Walk Optimisation Algorithm (QWOA) \cite{Marsh2019,marsh2020combinatorial} was designed to identify optimal, or near-optimal, solutions to combinatorial optimisation problems. Formally, we consider a mapping $f: \mathbb{S} \longrightarrow \mathbb{R}$, which returns a measure of the cost (or `quality') associated with each possible solution in the solution space $\mathbb{S}$, where $\mathbb{S}$ has cardinality $M$. 
    
    The starting point of the QWOA is a quantum system with $M$ basis states, one for each solution in $\mathbb{S}$, initialised in an equal superposition,
    
    \begin{equation}
        \ket{s} = \frac{1}{\sqrt{M}}\sum_{x \in \mathbb{S}}\ket{x}.
        \label{eq:eqsuperpos}
    \end{equation}
    
    \noindent This initial state is then evolved through repeated application of the \emph{quality-dependent phase-shift} and \emph{quantum-walk-mixing} unitaries. The quality-dependent phase-shift unitary
    
    \begin{equation}
        U_Q(\gamma_j) = \exp(-\text{i} \gamma_j Q ),
    \end{equation}
    
    \noindent where $\gamma_j \in \mathbb{R}$ and $Q$ is a diagonal operator such that $Q\ket{x} = f(x) \ket{x}$. The quantum-walk-mixing unitary is defined as
    
    \begin{equation}
        U_W(t_j) = \exp(-\text{i} t_j \mathcal{L} ),
    \end{equation}
    
    \noindent where $t_j \geq 0$, and $\mathcal{L}$ is the Laplacian matrix of a circulant graph that connects the feasible solutions to the problem. For the purpose of this work, we choose a complete graph that connects computational basis states that correspond to a valid solution to the problem. Specifically, the dimension of the Hilbert space is always a power of two whilst the total number of valid routing solutions is not, in general, a power of two. In order to ensure a valid solution to the problem, it is necessary to restrict the connectivity to only the computational basis states that correspond to valid solutions to the problem. Thus, the Laplacian is defined as
    
    \begin{equation}
    \bra{x} \mathcal{L} \ket{y} = \begin{dcases}
            0 & \text{$x$ or $y$ are not solutions} \, , \\
            M-1, & x=y \, , \\
            -1, & x \neq y \, .
        \end{dcases}
        \label{eq:lap}
    \end{equation}
    
    Note that the initial state, $\ket{s}$, can be understood as the complete graph connecting all solutions in the solution space, where each solution/node is occupied with equal probability. The first unitary $U_Q$ applies a phase-shift at each node proportional to the cost/quality of the solution at that node, with the proportionality constant given by the parameter, $\gamma_j$. The second unitary $U_W$ can be understood as performing a quantum walk over the complete graph for time $t_j$, mixing the amplitudes across nodes. Following the mixing of phase-shifted amplitudes across the nodes of the complete graph, constructive and destructive interference will result in quality-dependent amplitude amplification, controlled by the parameters $\gamma_j$ and $t_j$. Application of $U_Q$ and $U_W$ is repeated $r$ times such that the final state of the system is given by,
    
    \begin{equation} \label{eq:QWOA}
        \ket{\bm{\gamma}, \bm{t}}=U_{QWOA}(\bm{\gamma}, \bm{t})\ket{s} = U_W(t_r) U_Q(\gamma_r)...U_W(t_{1})U_Q(\gamma_{1}) \ket{s},
    \end{equation}
    
    \noindent where $\bm{t} = (t_1, t_2, ..., t_r)$ and $\bm{\gamma} = (\gamma_1, \gamma_2, ..., \gamma_r)$. 
    
     By tuning the parameters $\bm{\gamma}$ and $\bm{t}$, it is possible to amplify the amplitudes corresponding with low cost solutions, and therefore increase the probability of a measurement of the system collapsing it into a low cost solution. The process of tuning the parameters is conducted iteratively through the use of a classical optimisation algorithm (e.g. Nelder-Mead) which takes as its objective function the expectation value of the $Q$ operator:
    \begin{equation}
    \label{eq:expectation}
        c(\bm{\gamma}, \bm{t}) = \bra{\bm{\gamma}, \bm{t}}Q\ket{\bm{\gamma}, \bm{t}}.
    \end{equation}
    
    \noindent The QWOA is therefore a hybrid quantum (amplitude amplification) and classical (variational) approach. The extent of amplitude amplification possible is restricted by the number of iterations, so an increasing $r$ presents the opportunity for greater amplification of the amplitudes corresponding to the optimal or sub-optimal solutions in the final state $\ket{\bm{\gamma}, \bm{t}}$.
    
    \subsection{Quantum Circuits}
    
\begin{figure}
        \centering
        \includegraphics[width=0.95\columnwidth]{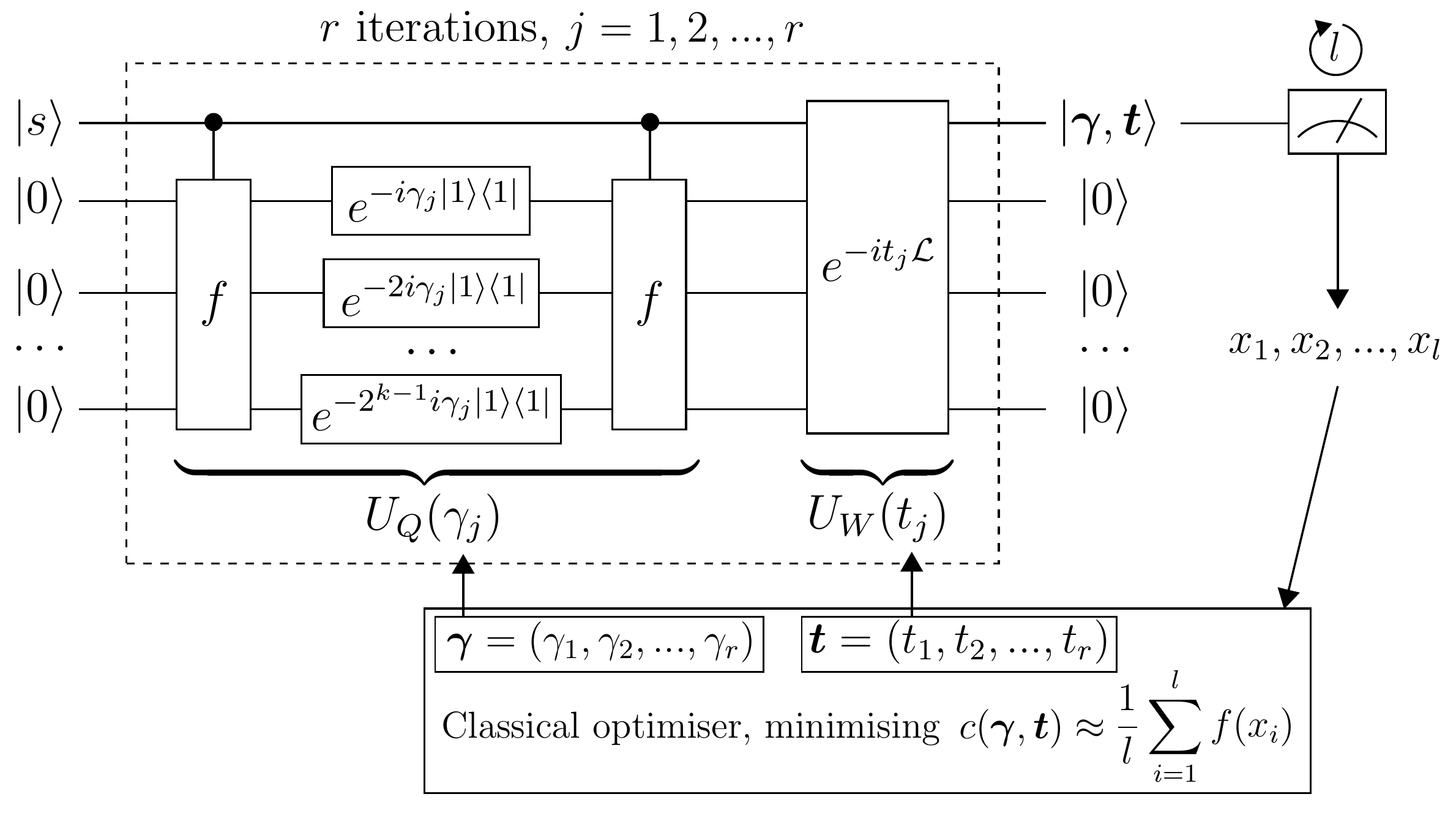}
        \caption{Schematic diagram of the QWOA circuit paired with a classical optimiser. The classical optimiser is used to tune the phase-shift and walk-time parameters, $\bm{\gamma}$ and $\bm{t}$, in order to produce an optimally amplified state, $\ket{\bm{\gamma}, \bm{t}}$, with a low expectation value for cost $c(\bm{\gamma}, \bm{t})$. The circuit performs $\ket{\bm{\gamma}, \bm{t}}\ket{0} = U_W(t_r) U_Q(\gamma_r)...U_W(t_{1})U_Q(\gamma_{1}) \ket{s}\ket{0}$. The graph Laplacian $\mathcal{L}$ encodes the connectivity between feasible solutions. The operator ${f}$ evaluates the quality of a solution to $k$ bits of precision and implements the costing algorithm presented in \cref{sec:algorithms}.}
        \label{fig:qaoa_circuit}
    \end{figure}

    \cref{fig:qaoa_circuit} illustrates the overall quantum circuit layout for the QWOA algorithm, with each iteration applying first the \emph{quality-dependent phase-shift} and then the \emph{quantum-walk-mixing} unitary. 
    %The circuit involves application of the indexing unitary $U_\#$, and uses the Laplacian of the complete graph $\mathbb{K}_M$ to connect solutions. 
    This circuit will complete $r$ iterations for a depth-$r$ QWOA, leading to $2r$ independent classical variational parameters. The expectation value of the solution quality is used to tune the parameters and is obtained by repeated sampling of the output state. This sampling process is efficient in obtaining a precise estimation of the expectation value for any problem in the NPO-PB complexity class as discussed in \cite{Marsh2019}.

%     \begin{figure}
% 	\centering
% 	\[ \Qcircuit @C=.9em @R=.5em {
% 		&&&&\push{\rule{2em}{0em}}&&\ustick{{U}_f(\gamma)}&&&&&\ustick{{U}_w(t)}&&\\
% 		&&&\lstick{\ket{\psi}} & \ctrl{1} & \qw & \qw & \qw & \ctrl{1} & \qw & \qw & \multigate{5}{e^{-i t \mathcal{L}}} & \qw & \qw & {\ket{\psi'}}\\
% 		&&&\lstick{\ket{0}} & \multigate{3}{{f}} & \qw & \gate{e^{-i \gamma \ket{1} \bra{1}}} & \qw & \multigate{3}{{f}} & \qw & \qw & \ghost{e^{-i t \mathcal{L}}} & \qw & \qw & {\ket{0}}\\
% 		\lstick{\rotatebox{90}{$\mathcal{O}(\text{polylog\,} M)$}} &&&\lstick{\ket{0}} & \ghost{{f}} & \qw & \gate{e^{-2i \gamma \ket{1} \bra{1}}} & \qw & \ghost{{f}} & \qw & \qw &\ghost{e^{-i t \mathcal{L}}} & \qw & \qw & {\ket{0}}\\
% 		&&&\lstick{\cdots} & & & \cdots & & & & & &  & &  \cdots\\
% 		&&&\lstick{\ket{0}} & \ghost{{f}} & \qw & \gate{e^{-2^{k-1} i \gamma \ket{1} \bra{1}}} & \qw & \ghost{{f}} & \qw & \qw  & \qw & \qw & \qw & {\ket{0}} \\
% 		&&& &  & &  & & & &  & &  & &
% 		{\gategroup{1}{5}{7}{10}{.7em}{--}\gategroup{3}{2}{6}{2}{.7em}{\{}\gategroup{1}{11}{7}{13}{.7em}{--}}
% 	}\]
% 	\caption[Generic quantum circuit for a QWOA iteration]{A single iteration of the QWOA circuit. The circuit performs $\ket{\psi'}\ket{0} = {U}_w(t) {U}_f(\gamma) \ket{\psi}\ket{0}$ with $\mathcal{O}(\text{polylog} \, M)$ depth. The graph Laplacian $\mathcal{L}$ encodes the connectivity between feasible solutions. The operator ${f}$ evaluates the quality of a solution to $k$ bits of precision. The rotation angles $\gamma$ and $t$ can be varied by a classical optimiser.}
% 	\label{fig:qaoa_circuit}
%     \end{figure}
    
    The QWOA utilises a quantum walk over the feasible combinatorial solutions, connected using an arbitrary choice of graph. The requirement on the graph is that there exists an efficient quantum circuit to simulate the quantum walk. We choose to connect the $M$ valid solutions using the complete graph $\mathbb{K}_M$, where $M$ is not necessarily a power of 2. The arbitrary-modulus Quantum Fourier Transform (QFT) was shown to be an efficient approach to perform quantum walk on this graph, and further was applicable to any other choice of circulant graph \cite{marsh2020combinatorial}. However, the QFT is known to be sensitive to noise \cite{ShorErrorCorrection}, which is an undesirable property for NISQ computation. In terms of the QWOA this implies potential degradation of the quantum walk, causing leakage of amplitude to infeasible solutions. Here, we present an improved highly efficient quantum circuit for fast-forwarded quantum walk simulation on complete graphs connecting the feasible solutions.

The Laplacian described in \cref{eq:lap} can be expressed as $\mathcal{L}=M(\mathbb{I}-\ket{s}\bra{s})$, where $\ket{s}$ is the equal superposition over the feasible solutions as per \cref{eq:eqsuperpos}. 
Thus the quantum walk can be expressed as, up to a global phase,
\begin{equation}
	e^{-i t \mathcal{L}} = \mathbb{I} + (-1 + e^{i M t}) \ket{s} \bra{s} \, .
\end{equation}
This expresses the rotation of the system about a specific state $\ket{s}$. The rotation about a specific state is well-known in the context of deterministic quantum search and optimal fixed-point amplitude amplification \cite{Long2001,Brassard2002,Yoder2014}. We can therefore directly utilise these results to design an exact circuit for a quantum walk on the complete graph (which connects only the feasible solutions). 
The first step is to prepare an equal superposition state over all feasible solutions. We do this by preparing the equal superposition over the first $M$ integers $(1/\sqrt{M})\sum_{y=0}^{M-1} \ket{y}$, and then use a specific \textit{un-indexing} unitary $U_{\#}^\dag$ to map each unique index state $\ket{i}$ to the $i$th feasible solution state $\ket{id^{-1}(y)}$, thus producing the equal superposition over the feasible solution states $\ket{s}$. 

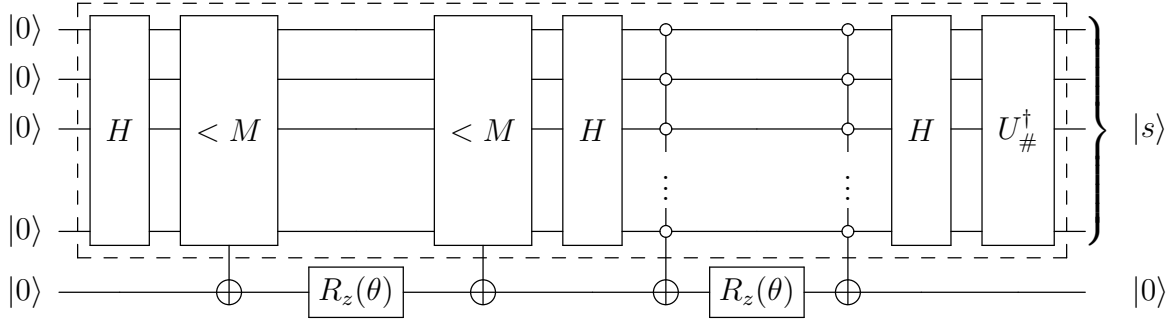
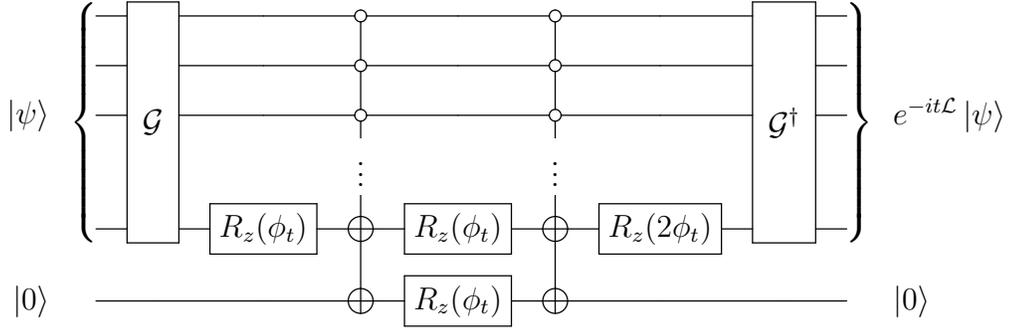
\begin{figure}
	\begin{subfigure}[t]{\linewidth}
		\centering
		\[ \Qcircuit @C=1em @R=.7em {
			\lstick{\ket{0}} & \multigate{4}{H}& \multigate{4}{<M} & \qw & \multigate{4}{<M} & \multigate{4}{H} & \ctrlo{1} & \qw& \ctrlo{1} & \multigate{4}{H} & \multigate{4}{U_{\#}^\dag} & \qw\\
			\lstick{\ket{0}} & \ghost{H}  & \ghost{<M}& \qw & \ghost{<M} & \ghost{H} & \ctrlo{1} & \qw& \ctrlo{1}& \ghost{H}& \ghost{U_{\#}^\dag} & \qw \\
			\lstick{\ket{0}} & \ghost{H} & \ghost{<M}& \qw & \ghost{<M}  & \ghost{H} & \ctrlo{2}& \qw& \ctrlo{2} & \ghost{H}& \ghost{U_{\#}^\dag} & \qw & \rstick{\ket{s}} \\
			 & \push{\rule{0em}{1em}} & & & & & \colorbox{white}{\vdots} & & \colorbox{white}{\vdots} \\
			\lstick{\ket{0}} & \ghost{H} & \ghost{<M} & \qw & \ghost{<M} & \ghost{H} & \ctrlo{1} & \qw& \ctrlo{1}& \ghost{H} & \ghost{U_{\#}^\dag}& \qw \\
			\lstick{\ket{0}}  & \push{\rule{0em}{1.5em}}\qw & \targ\qwx[-1] & \gate{R_z(\theta)} &  \targ\qwx[-1] & \qw & \targ & \gate{R_z(\theta)} & \targ & \qw & \qw & \qw & \rstick{\ket{0}}
			\gategroup{1}{2}{5}{11}{.8em}{--}
			\gategroup{1}{12}{5}{12}{.8em}{\}}
		}\]
		\caption{}
	\end{subfigure}
	\begin{subfigure}[t]{\linewidth}
		\centering
		\[ \Qcircuit @C=1em @R=.7em {
			& \multigate{4}{\mathcal{G}} & \qw & \ctrlo{1} & \qw & \ctrlo{1} & \qw & \multigate{4}{\mathcal{G^\dag}} & \qw \\
			& \ghost{\mathcal{G}}  & \qw & \ctrlo{1}& \qw & \ctrlo{1} & \qw & \ghost{\mathcal{G^\dag}} & \qw\\
			\lstick{\ket{\psi}\hspace{1em}} & \ghost{\mathcal{G}} & \qw & \ctrlo{2}& \qw & \ctrlo{2} & \qw & \ghost{\mathcal{G^\dag}} & \qw & \rstick{e^{-i t \mathcal{L}}\ket{\psi}}\\
			& \push{\rule{0em}{1em}} & & \colorbox{white}{\vdots} & & \colorbox{white}{\vdots}\\
			& \ghost{\mathcal{G}} & \gate{R_z(\phi_t)} & \targ & \gate{R_z(\phi_t)} & \targ & \gate{R_z(2\phi_t)} & \ghost{\mathcal{G^\dag}} & \qw \\
			\lstick{\ket{0}\hspace{1em}} & \qw  & \qw &  \targ\qwx[-1] & \gate{R_z(\phi_t)} &  \targ\qwx[-1] & \qw & \qw  & \qw & \rstick{\ket{0}}
			\gategroup{1}{1}{5}{1}{.8em}{\{}
			\gategroup{1}{9}{5}{9}{.8em}{\}}
		}\]
		\caption{}
	\end{subfigure}
	\caption{(a) Circuit for preparing the equal superposition over feasible solutions such that $\mathcal{G}\ket{0}\ket{0} = \ket{s}\ket{0}$, where $\theta = 2\arcsin\sqrt{\frac{2^m}{4M}}$ with $m=\ceil{\log_2 M}$. Here the $<M$ gate is an integer comparator that conditionally toggles the ancilla if the input register holds a value less than $M$. (b) Circuit for exact simulation of quantum walk on the complete graph connecting feasible solutions, where $\phi_t = -\frac{M t}{2}$, by rotating about the $\ket{s}$ state.}
	\label{fig:exactwalkcomplete}
\end{figure}

We label this unitary to prepare the desired state $\ket{s}$ starting in the $\ket{0}$ state as $\mathcal{G}$, as shown in \cref{fig:exactwalkcomplete}. A similar approach was used by \citet{Chiew2019} for preparing superpositions over set permutations. Let $m={\ceil{\log_2 M}}$ be the number of qubits, and $2^m$ the dimension of the corresponding Hilbert space. Clearly if $M$ is a power of $2$, we can set $\mathcal{G}=H^{\otimes n}$. Otherwise, define
\begin{equation}
	\mathcal{G} = U_{\#}^{\dag} H^{\otimes m} S_0(\theta) H^{\otimes m} S_\chi(\theta) H^{\otimes m} \, ,
\end{equation}
where
\begin{equation}
	S_0(\theta) = \begin{cases}
		e^{i \theta} \ket{j} & j = 0 \, , \\
		\ket{j} & 1 \leq j < 2^m \, ,
	\end{cases}
\end{equation}
and
\begin{equation}
	S_\chi(\theta) = \begin{cases}
		e^{i \theta} \ket{j} & 0 \leq j < M \, , \\
		\ket{j} & M \leq j < 2^m \, .
	\end{cases}
\end{equation}
It can be directly verified that by choosing $\theta = 2\arcsin\sqrt{\frac{2^m}{4M}}$, we have $\ket{s} = \mathcal{G} \ket{0}$ as required. Thus, utilising the fixed-point search quantum circuit given in \cite{Yoder2014}, we give a quantum circuit in \cref{fig:exactwalkcomplete} that exactly simulates quantum walk on the complete graph connecting the feasible solutions. 

Note that the spatial complexity of this QWOA circuit is dominated by that of the input register, which will require $m={\ceil{\log_2 M}}$ qubits, where M, the cardinality of the solution space, is given in \cref{sec:cardinality}. The time complexity of the quantum walk section of the circuit (besides the indexing process) is dominated by the use of the quantum comparator, leading to a gate complexity of $\mathcal{O}(m \log m)$ \cite{gidney2018factoring}. However, the indexing and un-indexing algorithms will dominate the overall circuit complexity, with a gate complexity of $\mathcal{O}(poly(n))$. Note that these algorithms are discussed in \cref{sec:indexing} and have classical time complexity of $\mathcal{O}(n^2)$. It is assumed that a quantum circuit implementation would therefor have gate complexity of at least $\mathcal{O}(n^2)$ but not greater than $\mathcal{O}(poly(n))$
    
 %   \subsection{Prerequisite Features for a QWOA Applicable Problem}

    \section{Prerequisite Features for QWOA}
    \label{sec:algorithms}
    
     There are four primary criteria for a problem to be considered as an adequate candidate for QWOA \cite{marsh2020combinatorial,slate2020quantum}. Specifically,
    \begin{enumerate}
    \item The cardinality of the solution space (number of unique solutions) $M$ must be efficiently computable, given a specific problem instance.
    \item There must exist an efficient indexing and un-indexing algorithm for the solution space. The indexing algorithm should be capable of taking any arbitrary solution and returning its location in an ordered list of the full solution space. The un-indexing algorithm should be able to do the reverse of this process, returning the unique solution corresponding to a given index. Both of these processes should be efficient, in other words, they can not simply involve the construction of the entire solution space and their complexity should be at most polynomial relative to the size of the problem. 
    \item The quality of any arbitrary solution must be efficiently computable.
    \item As per \cite{Marsh2019}, the problem must lie in the NPO-PB class (the class of optimisation problems where the objective function is bounded polynomially in the problem size). This requirement comes from sampling the quantum circuit to estimate the expectation value, to ensure the number of samples required does not grow exponentially.
    \end{enumerate}
The CVRP satisfies these criteria as discussed below. 
    
    \subsection{Cardinality of the Solution Space}
    \label{sec:cardinality}
    
    As described earlier, the full solution space for a problem of size $n$ can be represented by the set of all possible partitions of $n$ elements into nonempty and totally ordered subsets. This is closely related to the unsigned Lah numbers, $L(n,k)$, which count the number of ways a set of $n$ elements can be partitioned into $k$ nonempty and totally ordered subsets \cite{petkovvsek2007combinatorial}. It follows logically that the number of solutions, $M$, in the solution space for a problem of size $n$ would be given by the sum of unsigned Lah numbers:
    \begin{equation}
        M = L(n) = \sum_{k=1}^{n}L(n,k).
    \end{equation}
    
    \noindent The unsigned Lah Numbers can be computed with the  equation: 
    
    \begin{equation} \label{eq:lah_numbers}
        L(n,k)=  {n-1 \choose k-1} \frac{n!}{k!}. 
    \end{equation}
    
   \noindent Alternatively, the unsigned Lah numbers can be computed using the recursive relation, \cite{benyi2020generalized}:
    
    \begin{equation} \label{eq:lah_recursive}
        L(n,k)=
        \begin{dcases}
            1, & \text{if } n=k \\
            0, & \text{if } n,k=0 \text{ or } k>n \\
            L(n-1, k-1) + (n+k-1) L(n-1,k), & \text{otherwise}.
        \end{dcases}
    \end{equation}
    
    \noindent As such, the cardinality of the solution space for the CVRP is efficiently computable for a problem of any size.
    
    \subsection{Indexing and Unindexing Algorithms}
    \label{sec:indexing}
    
    The application of the QWOA to any particular problem requires that its solution space be efficiently indexable. The logic required to produce such an indexing algorithm for the solution space of a size $n$ CVRP can be found by closely analysing the recursive relation depicted in \cref{eq:lah_recursive}. 
    
    Solutions can  be classified according to $k$, the number of delivery groups (or subsets) that they contain, with $k$ taking values: $1,2,...,n$. As such, the solution space can be divided into $n$ distinct sub-spaces, one for each of these classifications.  Each of these sub-spaces can be further divided into those solutions where the $n^{\text{\tiny th}}$ element exists in a singleton, i.e. in a subset by itself, and those where it exists in a subset with other elements.
    
    For a given solution sub-space with $k = K$, the number of solutions contained is given by $L(n,K) = L(n-1, K-1) + (n+K-1) L(n-1,K)$ (see \cref{eq:lah_recursive}). With careful consideration, it can be seen that the first term in this expression corresponds to the group of solutions for which the $n^{\text{\tiny th}}$ element is in a singleton: by removing the $n^{\text{\tiny th}}$ element from every solution, a subset is lost, and the solution sub-space which remains consists of all ordered partitions of $n-1$ elements into $K-1$ subsets, of which there are $L(n-1, K-1)$. 
    
    Similarly, the second term corresponds to the group of solutions for which the $n^{\text{\tiny th}}$ element is not in a singleton. This group of solutions can be acquired by taking the solution sub-space consisting of all partitions of $n-1$ elements into $K$ ordered subsets, of which there are $L(n-1,K)$ solutions, and for each of these solutions, inserting the $n^{\text{\tiny th}}$ element into each of the $(n+K-1)$ available locations. 
    
    This is the logic upon which the algorithm is built, which consists of asking several questions for any particular solution: 
    
    \begin{itemize}
        \item How many subsets exist within the solution? 
        \item Is the $n^{\text{\tiny th}}$ element in a singleton?
        \item If not, in which of the $(n+K-1)$ available locations is it situated?
        \item After removing the $n^{\text{\tiny th}}$ element, where is the next element located?
    \end{itemize}
    
\begin{figure}[H]
        \centering
        \includegraphics[width=1\columnwidth]{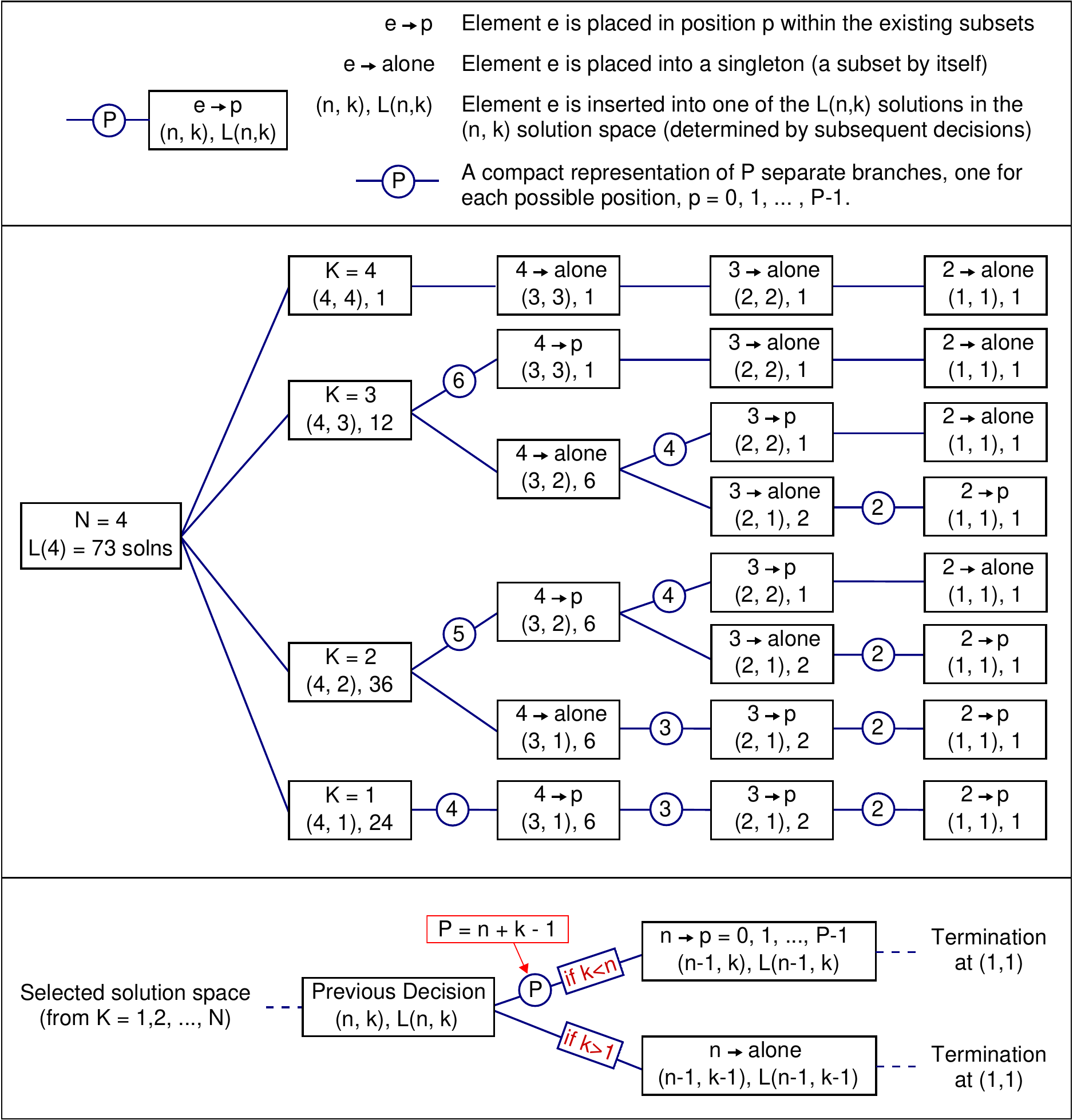}
        \caption{A compact decision tree for the CVRP of size $n=4$. The lower portion of the diagram shows the general structure and conditional logic which forms the basis for the construction of the tree. The branches labelled with a number represent multiple branches, with $p=0$ corresponding with the lower branch and so on.}
        \label{fig:decision_tree}
    \end{figure}

    \noindent The process of indexing or unindexing the solution space can also be understood as navigating a decision tree that is constructed from this same logic \cite{wilf1977unified}. An example decision tree for the problem size $n=4$ is shown in \cref{fig:decision_tree}.
    
    Each terminal node (those at the right-hand side in \cref{fig:decision_tree}) represents a solution in the solution space. The solution located at the bottom has index zero, and the indices increase incrementally up the tree, such that the index of any particular solution is equal to the number of solutions that exist below it. Note that the tree has been included in a compact form, though there appear to be only 8 terminal nodes, there are in fact 73. This is because the branches labelled with a number actually represent multiple branches, with the lowest branch being associated with $p=0$ and the branch above with $p=1$ and so on. Navigating through the decision tree from the origin to a particular terminal node involves choosing particular branches, each of which represents making decisions about the number of subsets in a solution, or the location of each element within the ordered subsets of a solution. By summing the number of solutions that are passed over with each decision made or branch taken, the resulting sum will be equal to the index of the final solution. In this way, the indexing process can be understood as navigating through the tree. Similarly, the unindexing process can be understood as choosing the appropriate branches such that the number of solutions passed over is equal to the index, with the terminal node corresponding to the required solution.
    
    The indexing and unindexing algorithms are shown in \cref{alg:Indexing} and \cref{alg:Unindexing}. The time complexities of both the indexing and unindexing algorithms scale with $n^2$, $\mathcal{O}(n^2)$, which satisfies the requirement that both of these algorithms should be efficiently computable. Since efficient classical algorithms for indexing and un-indexing exist, these algorithms can be directly `quantised' by translation to a reversible circuit. This defines the indexing unitary $U_{\#}$ used in \cref{fig:exactwalkcomplete}.
    
    \begin{algorithm}[H]
		\caption{Indexing: {This indexing algorithm returns the index of any solution, $soln$, from the solution space. It is composed of two parts, the first relates to placing the solution within a particular sub-space, determined by the number of subsets it contains, and the second is responsible for indexing the solution within this sub-space.
		As a clarification, on line \ref{alg:Indexing:line:p}, $p$ is assigned the location of the element $e$ within the solution/partition. This is the location where $e$ would have been placed in the partition not containing $e$, where the left-most location corresponds to $p=0$. This can also be calculated from the sum of the following: the number of subsets prior to the one containing $e$, the number of elements contained within these prior subsets, and the location of $e$ within its own subset.}}
		\label{alg:Indexing}
		\begin{algorithmic}[1]
    		\Function{index}{$soln$}    
                \State $n \gets \text{number of elements in } soln$ 
                \State $K \gets \text{number of subsets in } soln$
                \State \Return $\sum_{k=1}^{K-1}L(n,k) + \textsc{subindex}(n, K, soln)$
            \EndFunction
		    \State \Function{subindex}{$n,k,soln$}
    		    \State $e \gets \text{largest element in } soln$
    		    \If{$e=1$} \Return 0
    		    \ElsIf{$e$ \text{is in a singleton}}
    		        \State $\text{remove } e \text{ from } soln$
    		        \State \Return \textsc{subindex}$(n-1,k-1,soln)$
                \Else
    		        \State $p \gets \text{location of } e \text{ within } soln$ \label{alg:Indexing:line:p} 
    		        \State $\text{remove } e \text{ from } soln$
    		        \State \Return $L(n-1, k-1) + (n+k-1)\cdot\textsc{subindex}(n-1, k, soln)+p$
    		    \EndIf
    		\EndFunction
		\end{algorithmic}
	\end{algorithm}

    \begin{algorithm}[H]
		\caption{Unindexing: {This unindexing algorithm returns the solution to which a given index within the solution space of particular problem size, $n$, refers. The algorithm is composed of two  parts, the first is responsible for selecting the sub-space to which the indexed solution belongs, and its corresponding sub-index within that space and the second is responsible for returning the solution to which the sub-index and hence index refers.}}
		\label{alg:Unindexing}
		\begin{algorithmic}[1]
            \Function{unindex}{$n, index$}
                \State $subindex \gets index$
                \For{$k=1,2,\ldots,n$}
			        \If{$index < L(n,k)$} \Return $\textsc{solution}(n,k,subindex)$
			        \Else \text{ } $subindex \gets subindex - L(n,k)$
			        \EndIf
		        \EndFor
            \EndFunction
            \State \Function{solution}{$n,k,subindex$}
                \For{$e=n,n-1,\ldots,2$}
                    \If{$subindex < L(n-1, k-1)$}
                        \State element $e$ is placed in a singleton
                        \State $n \gets n-1$
                        \State $k \gets k-1$
                    \Else
                        \State $subindex \gets subindex - L(n-1, k-1)$
                        \State $p \gets subindex \bmod (n+k-1)$
                        \State element $e$ is placed in position $p$
                        \State $subindex \gets \left \lfloor subindex/(n+k-1) \right \rfloor$
                        \State $n \gets n-1$
                    \EndIf
                \EndFor
                \State construct $soln$ from the decisions made for each element, starting with a single set containing the element $1$
                \State \Return $soln$
            \EndFunction
        \end{algorithmic}
	\end{algorithm}

    \subsection{Computation of Solution Qualities}
    
    With the solution indexing now well established, the next challenge is to ensure the quality of an arbitrary solution can be efficiently computed and that the objective function is polynomially bounded in the number of locations. Computing the quality/cost corresponding to any given solution involves navigating the delivery network in the order specified by the solution while comparing the number of packages on the vehicle and its capacity with the number of packages required at each location/node. Each time an edge is traversed between two nodes, the corresponding cost from the cost matrix is incurred, and the total cost accumulated by the time the vehicle returns to the depot for the final time is the cost/quality of the solution. This process and the conditional logic involved is shown in detail in \cref{alg:computing_qualities_1} through ~\ref{alg:computing_qualities_3}. The complexity of this computation scales linearly with increasing problem size, $n$, so the quality of any arbitrary solution is efficiently computable, as required. In addition, for constant bounds on elements in the cost matrix, package vector, and the vehicle capacity, the maximum cost scales linearly with the number of locations and hence the objective function is sufficiently well-behaved for expectation sampling.
    
    \begin{algorithm}[H]
        \addtocounter{algsection}{+1}
        \caption{Computing Solution Qualities: {This algorithm returns the cost of a particular solution, $soln$, for a particular instance of the CVRP, defined by the parameters: vehicle capacity, $V$, package vector, $P$, and cost matrix, $C$. Note that the variable $LO$ is short for ``leftovers", and tracks the number of packages remaining on the vehicle. Also, the variable $perm$ is short for ``permutation", and $RH$ is short for ``return home". $RH$ is a Boolean variable that tracks whether the vehicle is due to return to the depot or has just returned to the depot.}}
        \label{alg:computing_qualities_1}
        \begin{algorithmic}[1]
            \Function{cost}{$soln, V, P, C$}
                \State $n \gets$ number of elements in $soln$
                \State $cost \gets 0$
                \For{$perm$ in $soln$}
                    \State $LO \gets V$
                    \State $RH \gets True$
                    \For{all $i$ (except the last) in $perm$}
                        \State $cost, LO, RH \gets \textsc{assess\_location}(i, cost, LO, RH, P, V, C, perm)$
                    \EndFor
                    \State $i \gets$ final element in $perm$
                    \State $cost \gets \textsc{assess\_final\_location}(i, cost, LO, RH, P, V, C)$
                \EndFor
                \State \Return $cost$
            \EndFunction
            \algstore{algorithm_3}
            \end{algorithmic}
            \end{algorithm}
             
            \begin{algorithm}[H]
            \addtocounter{algorithm}{-1}
            \addtocounter{algsection}{+1}
            \caption{Computing Solution Qualities}
            \label{alg:computing_qualities_2}
            \begin{algorithmic}
            \algrestore{algorithm_3}
            \Function{assess\_location}{$i, cost, LO, RH, P, V, C, perm$}
                \State $packages \gets P_i$
                \If{$RH = True$} $cost \gets cost + C_{0i}$
                \EndIf
                \If{$LO > packages$}
                    \State $LO \gets LO - packages$
                    \State $restocks \gets 0$
                    \State $RH = False$
                \Else 
                    \State $packages \gets packages - LO$
                    \If{$packages \mod V = 0$}
                        \State $restocks \gets \left \lfloor packages/V \right \rfloor $
                        \State $RH \gets True$
                        \State $LO \gets V$
                    \Else
                        \State $restocks \gets \left \lfloor packages/V \right \rfloor + 1$
                        \State $RH \gets False$
                        \State $LO \gets V - (packages\mod V)$
                    \EndIf
                \EndIf
                \State $cost \gets cost + restocks \cdot ( C_{0i} + C_{i0})$
                \If{$RH = True$} 
                    \State $cost \gets cost + C_{i0}$
                \Else
                    \State $j \gets$ next element in $perm$
                    \State $cost \gets cost + C_{ij}$
                \EndIf
                \State \Return $R, LO, RH$
            \EndFunction
            \algstore{algorithm_3}
            \end{algorithmic}
            \end{algorithm}
            
            \begin{algorithm}[H]
            \addtocounter{algorithm}{-1}
            \addtocounter{algsection}{+1}
            \caption{Computing Solution Qualities}
            \label{alg:computing_qualities_3}
            \begin{algorithmic}[1]
            \algrestore{algorithm_3}
            \Function{assess\_final\_location}{$i, cost, LO, RH, P, V, C$}
                \State $packages \gets P_i$
                \IIf{$RH = True$} $cost \gets cost + C_{0i}$
                \If{$LO > packages$} 
                \State $restocks \gets 0$
                \Else
                    \State $packages \gets packages - LO$
                    \State $restocks \gets \left \lfloor (packages-1)/V \right \rfloor + 1$
                \EndIf
                \State $cost \gets cost + restocks \cdot C_{0i}$
                \State $cost \gets cost + (restocks + 1) \cdot C_{i0}$
                \State \Return $cost$
            \EndFunction
            
        \end{algorithmic}
    \setcounter{algsection}{0}
    \end{algorithm}
    
    \section{Numerical Results}

    An example CVRP of size $n=8$ was created from randomly generated values and is captured by the following paramaters: \[ P = \left[ \begin{array}{c} 23 \\ 18 \\ 28 \\ 7 \\ 23 \\ 27 \\ 9 \\ 22 \end{array} \right], \;\; C= \left[ \begin{array}{ccccccccc} 0 & 10 & 16 & 10 & 14 & 17 & 12 & 11 & 17 \\ 10 & 0 & 7 & 8 & 14 & 9 & 4 & 1 & 5 \\ 16 & 7 & 0 & 15 & 10 & 10 & 5 & 2 & 11 \\ 10 & 8 & 15 & 0 & 5 & 15 & 13 & 15 & 15 \\ 14 & 14 & 10 & 5 & 0 & 1 & 4 & 15 & 4 \\ 17 & 9 & 10 & 15 & 1 & 0 & 13 & 5 & 3 \\ 12 & 4 & 5 & 13 & 4 & 13 & 0 & 2 & 7 \\ 11 & 1 & 2 & 15 & 15 & 5 & 2 & 0 & 2 \\ 17 & 5 & 11 & 15 & 4 & 3 & 7 & 2 & 0 \\ \end{array} \right] , \;\; V = 20 \]
    The inter-location costs consist of randomly generated integers from the uniform distribution from 1 to 15. Similarly, the depot to location costs are integers between 10 and 20, and the package counts between 5 and 30.
    
\begin{figure}[H]
        \centering
        \begin{subfigure}{1\columnwidth}
            \centering
            \includegraphics[width=1\columnwidth]{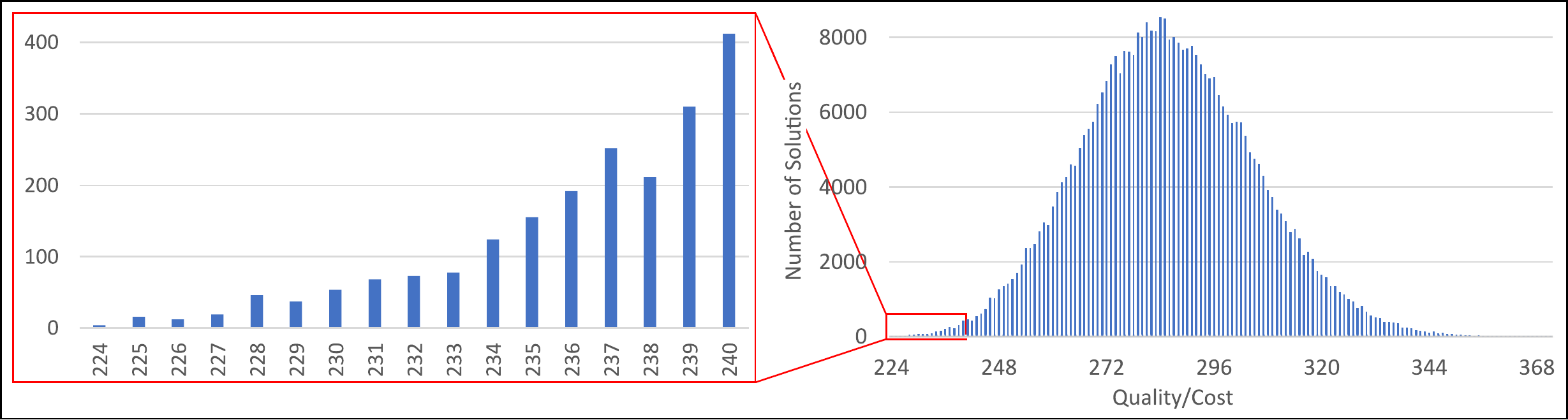}
            \caption{}
            \label{fig:initial_qual_dist}
        \end{subfigure}
        \begin{subfigure}{1\columnwidth}
            \centering
            \includegraphics[width=1\columnwidth]{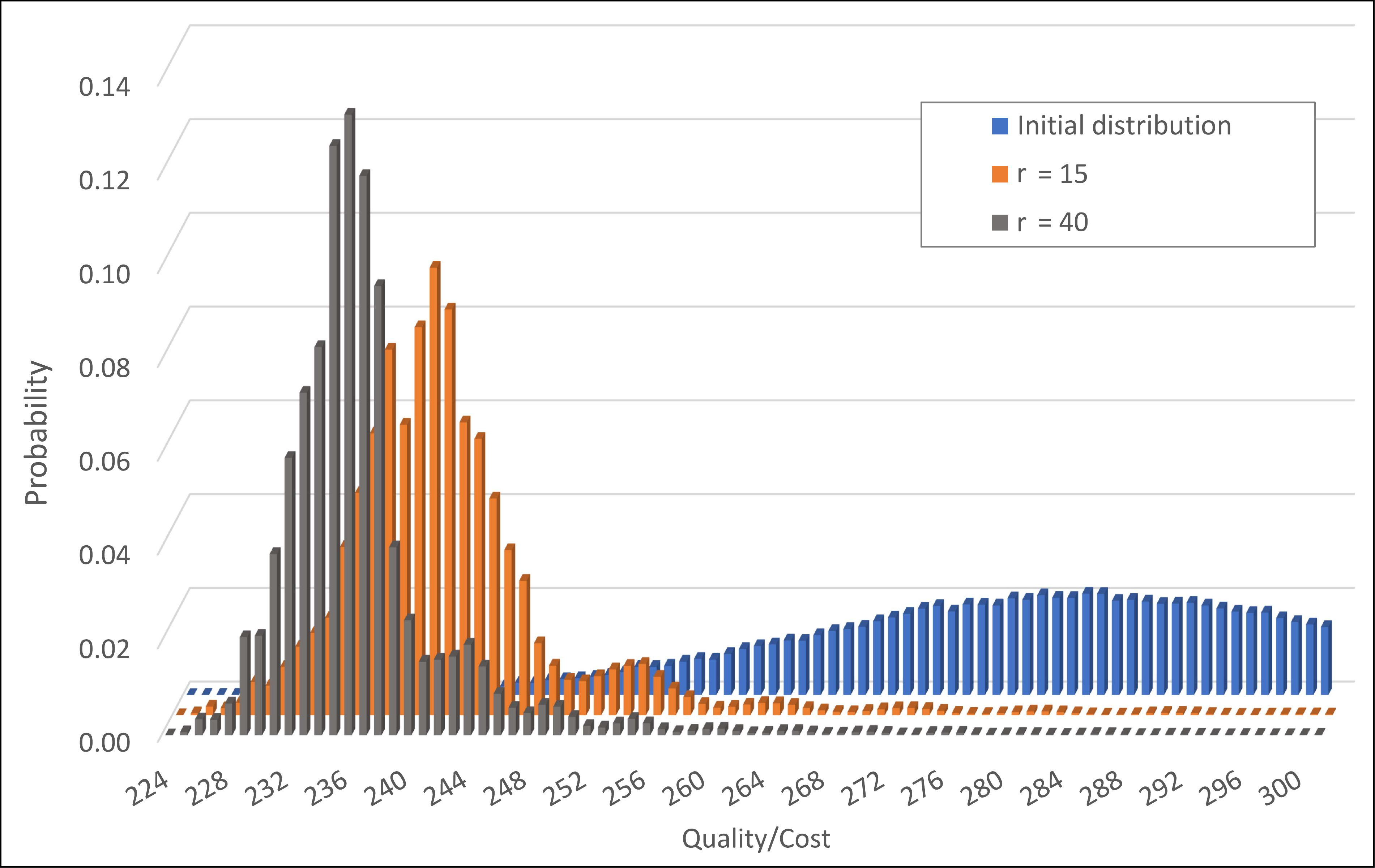}
            \caption{}
            \label{fig:probability}
        \end{subfigure}
        \caption{(a) Initial quality distribution of the example problem-solution space and, (b), the evolution of the probability distribution, relative to solution quality, with increasing iterations, $r$, of phase-shifts and walks.}
        \label{fig:quality_distributions}
    \end{figure}

    The cardinality of the solution space for an $n=8$ problem, calculated as per \cref{eq:lah_numbers}, is $L(8)=394,353$. Assessing the quality of these solutions reveals that there exist only 148 distinct solution qualities, where the lowest values correspond to the highest quality solutions. There is therefore a large amount of degeneracy in the qualities, which can be seen in the distribution of qualities shown in \cref{fig:initial_qual_dist}. 
    
    The distribution of qualities shown in \cref{fig:initial_qual_dist} is similar in nature to a normal distribution, though the distribution is skewed slightly towards the higher values (lower quality solutions). The solution space is populated predominantly by solutions of mid-tier quality, with only a small number of solutions of high quality.     
    
    In order to assess the capability of the QWOA to handle the example problem outlined above, the relevant quantum system and algorithm is simulated numerically. This is carried out using the software package, QuOP\_MPI \cite{quop_mpi_2020} in accordance with the framework laid out in Sec. IIIA. The phase-shift and walk-time parameters are optimised with the Broyden-Fletcher-Goldfarb-Shanno (BFGS) algorithm, implemented via the SciPy open-source Python library.   
    
    Shown in \cref{fig:initial_qual_dist} is the initial quality distribution of the example problem described above, prior to the application of any phase-shifts or quantum walks. Even though the initialisation of the system as an equal superposition means each node on the complete graph is associated with an equal part of the overall probability density, the mid-tier qualities are over-represented in the initial quality distribution because there are a large number of nodes corresponding to these qualities. The success of the QWOA will be measured by its ability to evolve the state of the system, via phase-shifts and quantum walks, such that the probability densities concentrate at the nodes corresponding with the highest quality solutions. 
    
\begin{figure}[H]
        \centering
        \includegraphics[width=0.85\columnwidth]{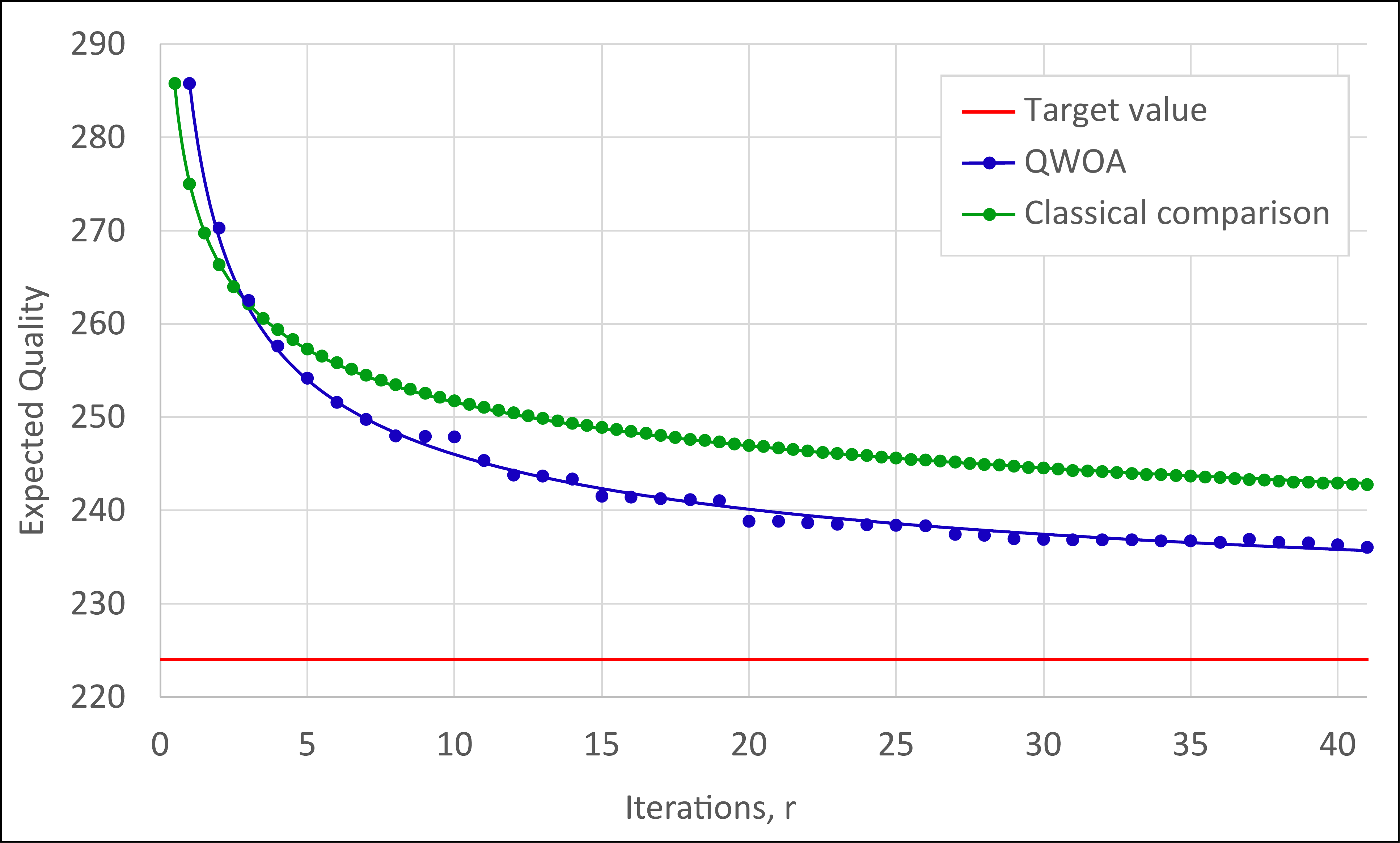}
        \caption{The convergence of the QWOA objective function towards the minimum/target value with increasing r as compared to a classical sampling method of equivalent computational effort. The QWOA curve scales as $1/r^{0.45}$, while the classical curve scales as $1/r^{0.27}$.}
        \label{fig:convergance_rate}
    \end{figure}

    The QWOA objective function is the expectation value of the measured solution quality after $r$ QWOA iterations, given by \cref{eq:expectation}. As shown in \cref{fig:probability}, with an increasing number of repeated iterations, $r$, the probability distribution within the system does in fact concentrate towards higher quality solutions. Upon closer inspection, however, for the given r values, the resulting quality distribution is not concentrating at the most optimal solutions as much as it is at the near-optimal solutions. With significantly larger r, it is expected that convergence towards the most optimal solutions would become more complete, but as shown in \cref{fig:convergance_rate}, the rate of convergence decreases at large r.  More interesting is a direct comparison with classical random sampling of equivalent computational effort, where the classical data represents the expected best quality measured from $2r$ classical random samples of the solution space, computed from 100,000 trials. Note that we compare with $2r$ classical samples, because this represents the same number of calls to the quality function when compared with $r$ iterations of the QWOA process, where we quantify computational effort by the number of calls to the quality function. It should also be clarified that no allowance has been made for the computational effort involved in the classical optimisation procedure to arrive at an optimal set of parameters, $\bm{\gamma}$ and $\bm{t}$, instead, the focus has been on the final optimally amplified state. This serves as a proof of concept in that it shows that the QWOA procedure is capable of providing speed up via sampling of the amplified state, given that there exists a computationally efficient method to arrive at a set of optimal parameters. \cref{fig:convergance_rate} shows that the expected quality/cost measured from a QWOA amplified state converges towards the target/minimum value as $\frac{1}{r^{0.45}}$, while the classical random sampling scales as $\frac{1}{r^{0.27}}$, when considering equivalent computational effort.

    \begin{figure}[H]
        \centering
        \includegraphics[width=0.7\columnwidth]{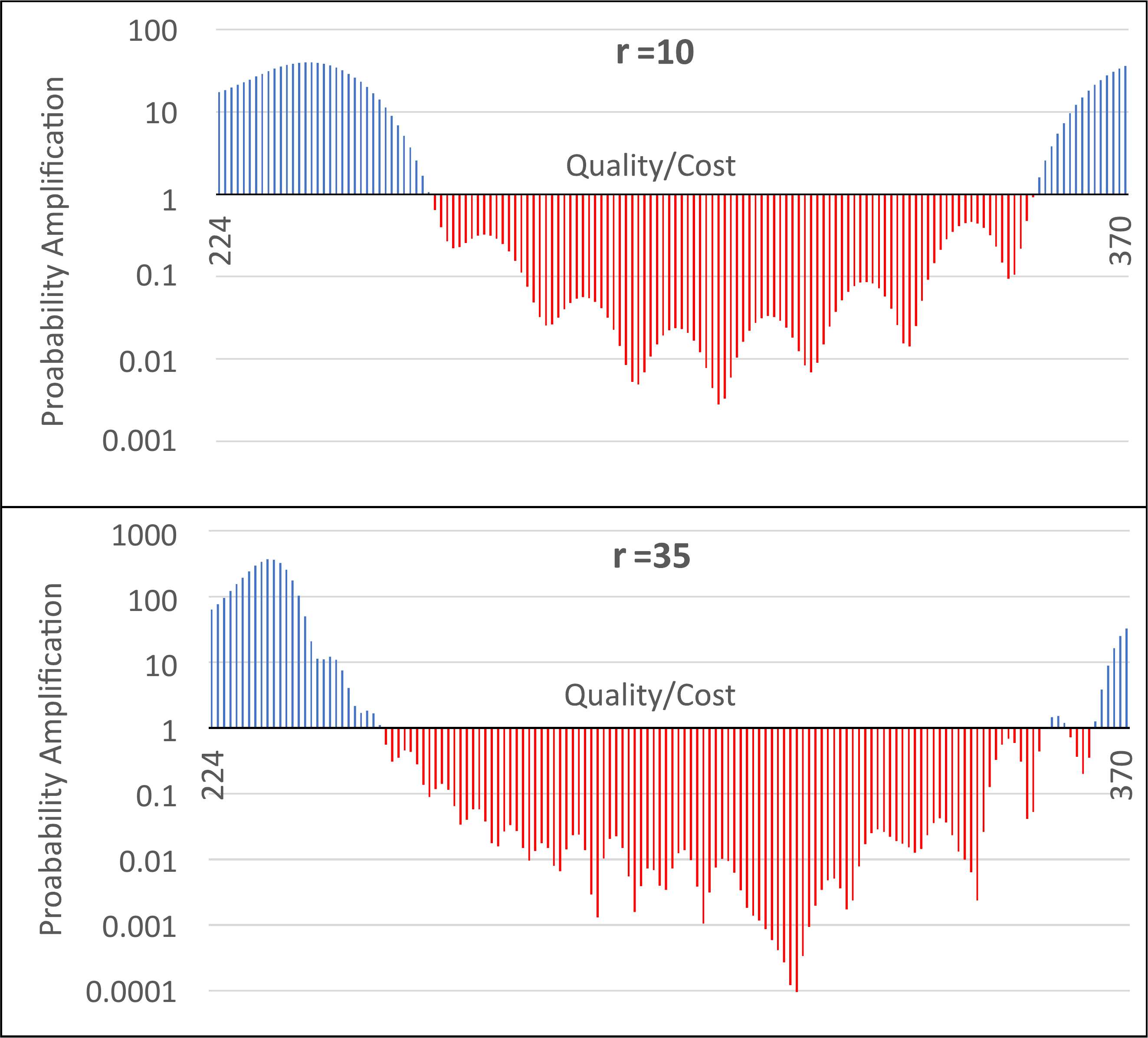}
        \caption{This graph quantifies the probability amplification at any particular node/solution as a function of its quality for the cases where $r=10$ and $r=35$}
        \label{fig:prob_amplification}
    \end{figure}
    
    Taking a closer look at how the probability densities are being amplified relative to quality may reveal part of the reason why the system seems to concentrate primarily towards sub-optimal solutions. The graph in \cref{fig:prob_amplification} shows how the probability densities are amplified at any particular node as a function of its solution quality for the $r=10$ and $r=35$ case. It is perhaps not surprising that the graph in \cref{fig:initial_qual_dist} shows more probability density at the near-optimal solutions, because there are significantly more of these solutions present in the solution space. The probability amplification shown in \cref{fig:prob_amplification}, on the other hand, is a comparison between final and initial probability density at any given node, so the number of solutions at a given quality does not have a multiplicative effect on the data shown. It is therefore quite telling that here the near-optimal solutions are being amplified more strongly than the most optimal solutions. A potential explanation for this is that amplification of near-optimal solutions is favoured by the optimisation process over the most optimal solutions, because although they are not as optimal, they have a larger influence in minimising the objective function due to their superior numbers. 
    
    Any way in which the most optimal solutions can be weighted more heavily with regards to their effect on the objective function would be beneficial in counteracting this potential bias towards sub-optimal solutions. However, this may not be effective in a practical sense, because the expectation value is calculated from repeated observations of the system, and in order for the increased weighting to take effect, the most-optimal solutions must actually be measured, which is not likely in the pre-amplified state.

    In the cannon of currently discovered quantum algorithms, QWOA is most naturally compared to the Quantum Approximation Optimisation Algorithm (QAOA). This algorithm seeks to solve combinatorial optimisation problems following a similar alternating ansatz that is capable of solving the CVRP problem as formulated in this study. However, QAOA is limited to a quantum walk across the entire Hilbert space \cite{slate2020quantum}, with invalid states being differentiated via a penalty function. In contrast, the QWOA indexing unitary restricts its equivalent walk operation to the subspace of only valid solutions. A previous numerical study demonstrated that this allows for the QWOA algorithm to produce more effective convergence to high-quality solutions than QAOA at the same circuit depth \cite{slate2020quantum}.
 
    As QWOA seeks approximate solutions to the CVRP combinatorial optimisation problem, it is comparable in intent to classical heuristic and metaheuristic algorithms. Such methods are numerous and, as with QWOA,  exhibit effectiveness that often depends on the structure of the underlying CVRP instance \cite{sanchez_comparative_2020}. Nevertheless, current methods can identify solutions within 0.5\% to 1\% of the optimum for problems with hundreds to thousands of delivery locations \cite{vidal_unified_2014}. The effectiveness of these algorithms is determined by benchmarking their performance against established data sets \cite{vidal_unified_2014,sanchez_comparative_2020}, which exceed the scale afforded by classical simulation of quantum dynamics. For these reasons, a comparison between this QWOA application and current classical heuristic and metaheuristic methods lies beyond the scope of this study.
    
    \section{Conclusion}
    
     To conclude, the quantum-walk-based optimization algorithm (QWOA) can be effectively applied to the Capacitated Vehicle Routing Problem. Efficient algorithms have been developed for the indexing and unindexing of the Solution space for this problem. It has been demonstrated for a randomly generated 8 location problem, that the QWOA is capable of significantly amplifying probabilities for optimal and near-optimal solutions, and in doing so, achieves expected qualities with less computational effort than that required by classical random sampling. The rate of convergence towards the most optimal solutions is however still somewhat limited, and the speed-up relative to classical sampling is contingent on navigating the variational process efficiently.  Further work will be focused on quantifying and minimising the computational effort involved in the variational process and in finding a way to improve the algorithm further to provide faster convergence towards the most optimal solutions.

    \section{Acknowledgements}
    
    This work was supported by resources provided by the Pawsey Supercomputing Centre with funding from the Australian Government and the Government of Western Australia. EM and SM acknowledge the support of the Australian Government Research Training Program Scholarship. SM's research was also supported by a Hackett Postgraduate Research Scholarship at the University of Western Australia. JBW would like to thank Shakib Daryanoosh for their useful discussions.
    
	\bibliography{refs}
	
\end{document}